\documentclass[a4paper,11pt]{article}
\usepackage{jheppub} 
\usepackage{lineno}
\usepackage{slashed}
\pdfoutput=1

\usepackage[normalem]{ulem}
\usepackage{siunitx}
\usepackage{subcaption}
\usepackage[font=small,labelfont=bf]{caption}
\usepackage{feynmp-auto}
\usepackage{placeins}

\DeclareSIUnit\year{yr} 

\title{\boldmath Electromagnetic multipoles of Majorana spin-3/2 dark matter particles and implications for direct detection}

\author[a]{Alejandro Ibarra}
\author[a]{Onur Yonar}
\affiliation[a]{Technical University of Munich, TUM School of Natural Sciences, Physics Department, 85748 Garching, Germany}
\emailAdd{ibarra@tum.de}
\emailAdd{onur.yonar@tum.de}

\abstract{We investigate the electromagnetic interactions of Majorana spin-3/2 dark matter and their implications for direct detection experiments. We derive model independent constraints on the three electromagnetic multipole moments of the spin-3/2 dark matter allowed by the gauge and Lorentz-invariances from the non-observation of dark matter induced nuclear recoils in direct detection experiments. We then calculate the expected values of the multipole moments in  two concrete particle physics scenarios: the gravitino in supergravity models, and a simplified model where a spin-3/2 particle interacts with an electrically charged spin-1/2 fermion and a vector boson.}

\begin{document}
\maketitle
\flushbottom
\section{Introduction}

The nature of the dark matter of the Universe remains one of the most important open questions in particle physics. Although astrophysical and cosmological observations strongly constrain the electric charge of dark matter to be very small~\cite{Dubovsky:2003yn,McDermott:2010pa,Dolgov:2013una}, dark matter may nevertheless interact with the electromagnetic field through higher-order multipole moments. Such interactions can arise at the quantum level from electrically charged particles coupled to the dark matter candidate, and may induce observable signals in direct and indirect detection experiments. The phenomenology of dark matter with electromagnetic multipole moments has been extensively investigated in the literature, both within an effective field theory approach and in specific particle physics models, {\it e.g.}~\cite{Pospelov_2000bq,Sigurdson:2004zp,Masso:2009mu,Fitzpatrick:2010br,Banks:2010eh,Barger:2010gv,Fortin:2011hv,Ho,DelNobile:2014eta,Kopp:2014tsa,Bai:2014osa,Garny:2015wea,Ibarra:2015fqa,Sandick:2016zut,Hambye:2021xvd,Tomar:2022ofh,Hisano:2020qkq,Ibarra_2023,Ibarra:2024mpq}.

The possible electromagnetic interactions of a particle are strongly constrained by its spin and by its transformation properties under discrete symmetries. In particular, Majorana spin-1/2 fermions cannot possess electric or magnetic dipole moments, and their only allowed diagonal electromagnetic form factor is the anapole~\cite{osti_4309791,Nieves:1981zt,
Kayser:1982br}. The corresponding interaction couples the axial current of the Majorana fermion to the electromagnetic current and can mediate elastic scattering with nuclei through the exchange of a virtual photon. This possibility has received considerable attention in the context of dark matter phenomenology, and the anapole moment generated at one loop has been calculated in several extensions of the Standard Model~\cite{Pospelov_2000bq,Ho,DelNobile:2014eta,Kopp:2014tsa,Bai:2014osa,Garny:2015wea,Ibarra:2015fqa,Sandick:2016zut}. Of particular interest is the neutralino in supersymmetric theories, whose anapole moment and implications for direct detection have been discussed in detail in~\cite{Ibarra_2023}.

The electromagnetic interactions of higher-spin particles are more intricate. In particular, the electromagnetic vertex of a Majorana spin-3/2 fermion has a richer Lorentz structure than its spin-1/2 counterpart, allowing additional form factors compatible with electromagnetic gauge invariance and the Majorana condition~\cite{Radescu1985,formfactor1,hypercharge, Nieves2013}. More recently, the phenomenology of higher-spin dark matter particles interacting through hypercharge anapole operators was investigated in~\cite{hypercharge}, including the implications for direct detection and collider searches. These studies demonstrate that electromagnetic interactions can provide a useful probe of higher-spin dark matter, although the size of the corresponding form factors ultimately depends on the underlying particle physics model.

In this work we investigate the electromagnetic interactions of Majorana spin-3/2 dark matter particles and their implications for spin-3/2 dark matter detection. We first adopt a model-independent approach, parametrizing the electromagnetic vertex in terms of three form factors. We calculate the differential cross section for elastic scattering with nuclei and derive constraints on the corresponding static moments from the non-observation of dark matter signals in direct detection experiments. We find that the different contributions exhibit distinct dependences on the momentum transfer, with the second and third form factors suffering additional kinematical suppressions at small nuclear recoil energies.

We then investigate the electromagnetic form factors in two concrete realizations of Majorana spin-3/2 dark matter. First, we calculate the electromagnetic form factors of the gravitino in the Minimal Supersymmetric Standard Model coupled to supergravity. We find that the non-vanishing one-loop contributions arise from diagrams involving chiral supermultiplets and that, in the static limit the $P$-violating anapole moments remain finite. The corresponding interactions are, however, strongly suppressed by the Planck scale. Second, we consider a simplified model where the dark matter particle interacts with an electrically charged vector boson and a charged fermion. We discuss in detail the dependence of the induced anapole moments on the particle masses and couplings, and compare the predictions with the sensitivities of direct detection experiments.

The paper is organized as follows. In section~\ref{sec:AnapoleGeneric} we discuss the general electromagnetic vertex of a Majorana spin-3/2 particle and derive constraints on its multipole moments from direct detection experiments. In section~\ref{sec:gravitino} we investigate the electromagnetic properties of the gravitino. In section~\ref{sec:simplified-model} we calculate the anapole moments induced at one loop in a simplified model with a charged vector boson and a charged fermion. Finally, in section~\ref{sec:conclusions} we present our conclusions.

\section{Scattering of spin-3/2 Majorana dark matter with nuclei through electromagnetic multipoles}
\label{sec:AnapoleGeneric}

We assume that the dark matter is constituted by a population of spin-3/2 Majorana fermions, described by vector spinors $\Psi_\mu$  with mass $m_{3/2}$, and Lagrangian with the Rarita-Schwinger form~\cite{RSF}:
\begin{equation}
    \mathcal{L}= -\dfrac{1}{2} \varepsilon^{\mu\nu\rho\sigma} \bar{\Psi}_\mu \gamma_5 \gamma_\nu \partial_\rho \Psi_\sigma - \frac14 m_{3/2} \bar{\Psi}_\mu \left[\gamma^\mu,\gamma^\nu\right] \Psi_\nu.
    \label{eq:RS}
\end{equation}

The Rarita-Schwinger field satisfies the Dirac equation together with the subsidiary conditions $\gamma^\mu\Psi_\mu=0$ and $\partial^\mu\Psi_\mu=0$. These constraints eliminate the unphysical spin-$1/2$ components of the vector-spinor field, leaving the four physical degrees of freedom of a massive spin-$3/2$ particle.

The Lorentz and gauge symmetries allow non-renormalizable interactions of the spin-3/2 Majorana fermion with the photon, described by an invariant amplitude of the form~\cite{Radescu1985,formfactor1,hypercharge}\footnote{In reference~\cite{Nieves2013} a more restricted form of the covariant vertex is presented. In this paper we only consider electromagnetic gauge invariance and on-shell particle constraints and assume no further relation among Lorentz generators. }
\begin{equation}\label{eq:current}
      M_{\rho}= \overline{\Psi}^{\nu}(l)   \Tilde{\Pi}_{\mu \nu \rho}(q) \Psi^{\mu}(p),
\end{equation}
where the covariant vertex is given by:
\begin{align} \label{eq:3}
       \Tilde{\Pi}_{\mu \nu \rho}(q) =  & A_1(q^2) \left(q^2 \gamma_{\rho} \gamma_5  -  q_{\rho}\slashed{q}\gamma_5 \right) g_{\mu \nu}\nonumber \\
       +& \left(  \gamma_5 A_2(q^2) + A_3(q^2)\right)\left(q_\mu q_\nu q_\rho -\dfrac{q^2}{2} ( g_{\rho \nu} q_\mu +  g_{\mu \rho} q_\nu) \right).
\end{align}
Here, $A_1(q^2)$, $A_2(q^2)$ and $A_3(q^2)$ are form factors; $A_1(q^2)$ and $A_2(q^2)$ are $C$-odd, $P$-odd and $CP$-even, while $A_3(q^2)$ is  $C$-odd, $P$-even and $CP$-odd. In the limit $q^2\rightarrow 0$ (approximately valid for dark matter direct detection experiments), one can define anapole moments of the spin-3/2 particle as the static limit of the form factors $a_i = \lim_{q^2\to 0} A_i(q^2)$, $i=1,2,3$.

The differential scattering cross section of the spin-3/2 dark matter $\Psi_\mu$ with a target nucleus $\text{N}_\text{T}$ with mass $m_N$, atomic number $Z$ and mass number $A$ is:
\begin{equation}\label{eq:sigma}
    \begin{aligned}
        \dfrac{d\sigma}{dE_R}=&\dfrac{5 e^2  }{36 \pi}\left( F_Z^2(q^2) Z^2 \left[2 m_N - \left(1 + \dfrac{m_N}{m_{3/2}}\right)^2 \dfrac{E_R}{v^2}\right]+ F_D^2(q^2)  \dfrac{2 m_N^2}{3 m_{p}^2} \left(\dfrac{\overline{\mu_N}}{\mu_p} \right)^2 \dfrac{E_R}{v^2}\right)\\  &\quad \times \bigg( |a_1|^2 + \dfrac{m_N^2 E_R^2}{4m_{3/2}^2}  |a_2|^2  + \dfrac{3 m_N E_R}{10} |a_3|^2
        + \dfrac{ m_N E_R}{5m_{3/2}}  \text{Re}\{a_1 a_2^{\dag}\} \bigg),
    \end{aligned}
\end{equation}
where $ F_Z^2(q^2)$ and $F_D^2(q^2)$ are the nuclear charge and magnetic dipole form factors
\begin{equation}
\begin{aligned}
     F_Z^2(q^2)&= \left(\dfrac{3 j_1(q R)}{q R}\right)^2 e^{-q^2 s^2},\\
     F_D^2(q^2)&= \begin{cases}
\left[ \dfrac{\sin(q R_D)}{q R_D} \right]^2, & q R_D < 2.55, q R_D > 4.5, \\[12pt]
0.047, & 2.55 \leq q R_D \leq 4.5, 
\end{cases}
\end{aligned}
\end{equation}
with $j_1(q R)$ a spherical Bessel function of the first kind and $R=\sqrt{c^2+7/3 \pi^2 a^2 -5s^2}$  the effective nuclear radius. Further, $c=(1.23 A^{1/3}-0.60)\hspace{0.05cm} \text{fm}$, $a=\SI{0.52}{\femto\meter}$, $R_D \simeq 1.0 A^{1/3}\hspace{0.05cm}\text{fm}$, and $s=\SI{0.9}{\femto\meter}$ is the nuclear skin thickness~\cite{FormfactorNuclear,helm11}. Further, $\mu_p = e /2 m_p$ denotes the nuclear magneton with the proton mass $m_p$, and $\overline{\mu_N}$ is the weighted dipole moment for the target nuclei, defined as: 
\begin{equation}
    \overline{\mu_N} = \left(\sum_i f_i \mu_i^2 \dfrac{S_i+1}{S_i} \right)^{1/2},
\end{equation}
where $f_i$, $\mu_i$, and $S_i$ are the elemental abundance, nuclear magnetic moment, and spin of the isotope $i$~\cite{cheng}.

We note that in Eq.~(\ref{eq:sigma}) the terms proportional to $A_3$ are suppressed by $E_R^{1/2}$ whereas terms proportional to $A_2$ have a kinematical suppression factor of $E_R$. The term in the square brackets is typical for fermions, as demonstrated in~\cite{hypercharge,Ho}, and it leads to an overall suppression of the differential scattering cross section $d\sigma/dE_R$.

The differential event rate at a direct detection experiment reads:
\begin{equation}\label{eq:ratemaster}
    \dfrac{dR}{dE_R}= \dfrac{\rho_{loc}}{m_N m_{3/2}}\int d^3v v f_{lab}(\Vec{v})  \dfrac{d\sigma}{dE_R}
\end{equation}
where $\rho_{loc}$ is the local density of dark matter, $m_N$ is the mass of the nucleus, $\Vec{v}$ is the relative velocity of the DM relative to Earth, $f_{lab}(\Vec{v})$ the distribution of velocities of the DM in the laboratory frame of the Earth and $d\sigma/dE_R$ gives the differential cross section per recoil energy for DM-nucleus scattering. The determination of the local DM density and velocity distribution is crucial for direct detection phenomenology. In our direct detection analysis we assume a single dark matter species with the dark matter density $\rho_{loc}=\SI{0.3}{\giga\electronvolt\per\centi\meter\cubed}$ \cite{LocalDensity}. 

Further, we assume that the DM velocity distribution follows a Maxwell-Boltzmann distribution in the galactic frame, truncated at the escape velocity from the DM halo, $v_{esc}$ \cite{Green:2011bv}:
\begin{equation}
    f_{lab}(\Vec{v})=f(\Vec{v}+\Vec{v_E})
\end{equation}
where $\Vec{v_E}$ is the velocity of the Earth in the galactic frame, and:
\begin{equation}  
      f(\Vec{v})= 
\begin{cases}
    \frac{1}{\mathcal{N}} e^{-v^2/v_0^2}&, \text{if } |\Vec{v}|< v_{esc}\\
    0              &, \text{otherwise}
\end{cases}
\end{equation}
with:
\begin{equation}
    \mathcal{N}= \pi^{3/2} v_0^3 \left[\text{erf}(\dfrac{v_{esc}}{v_0})-\dfrac{2 v_{esc}}{\sqrt{\pi}v_0} e^{-v_{esc}^2/v_0^2}  \right].
\end{equation}
Here, $\mathcal{N}$ is a normalization constant, $\text{erf($z$)}=(4/\pi)^{1/2}\int_0^z e^{-t^2} dt $ is the Gaussian error function and $v_0$ is the local circular velocity. For our analysis, we use the values $v_{esc}=\SI{544}{\kilo\meter\per\second}$, $v_0=\SI{220}{\kilo\meter\per\second}$ and $v_E=\SI{232}{\kilo\meter\per\second}$~
\cite{Smith:2006ym,Green:2011bv,FormfactorNuclear,Baxter:2021pqo}.

\begin{figure}[t!]
    \centering
        \includegraphics[width=.49\textwidth]{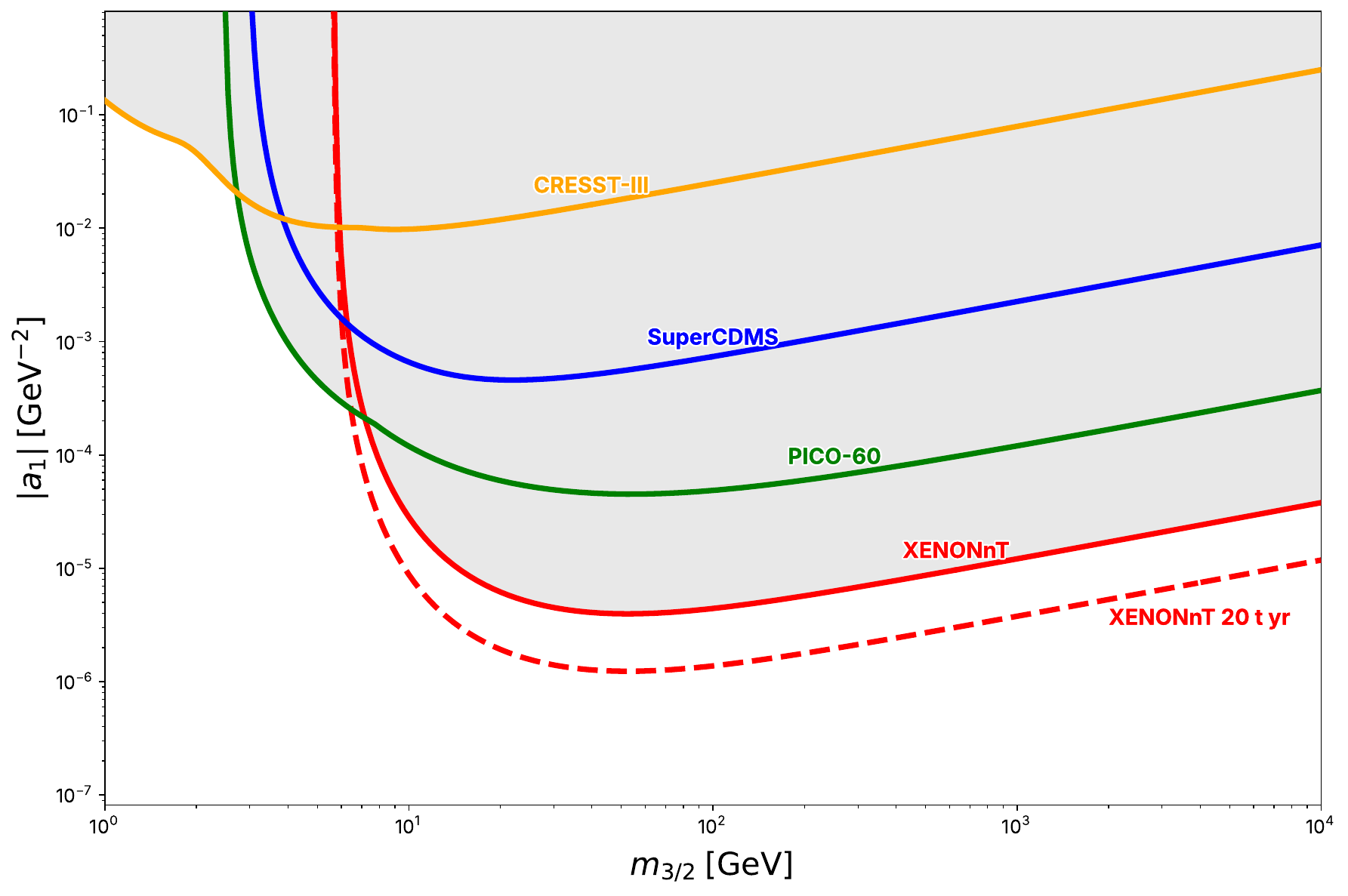}
        \includegraphics[width=0.49\textwidth]{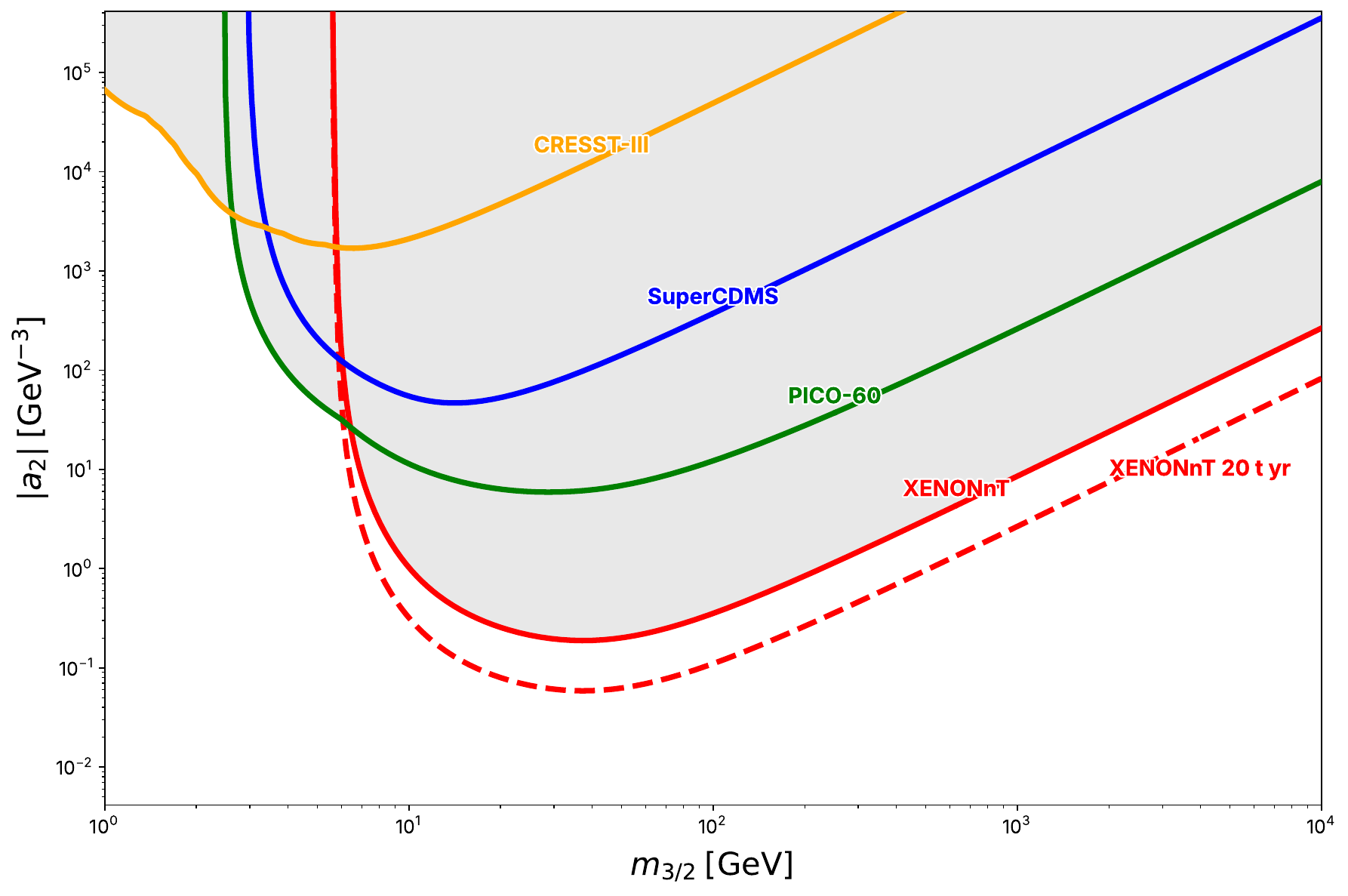}\\
        \includegraphics[width=0.49\textwidth]{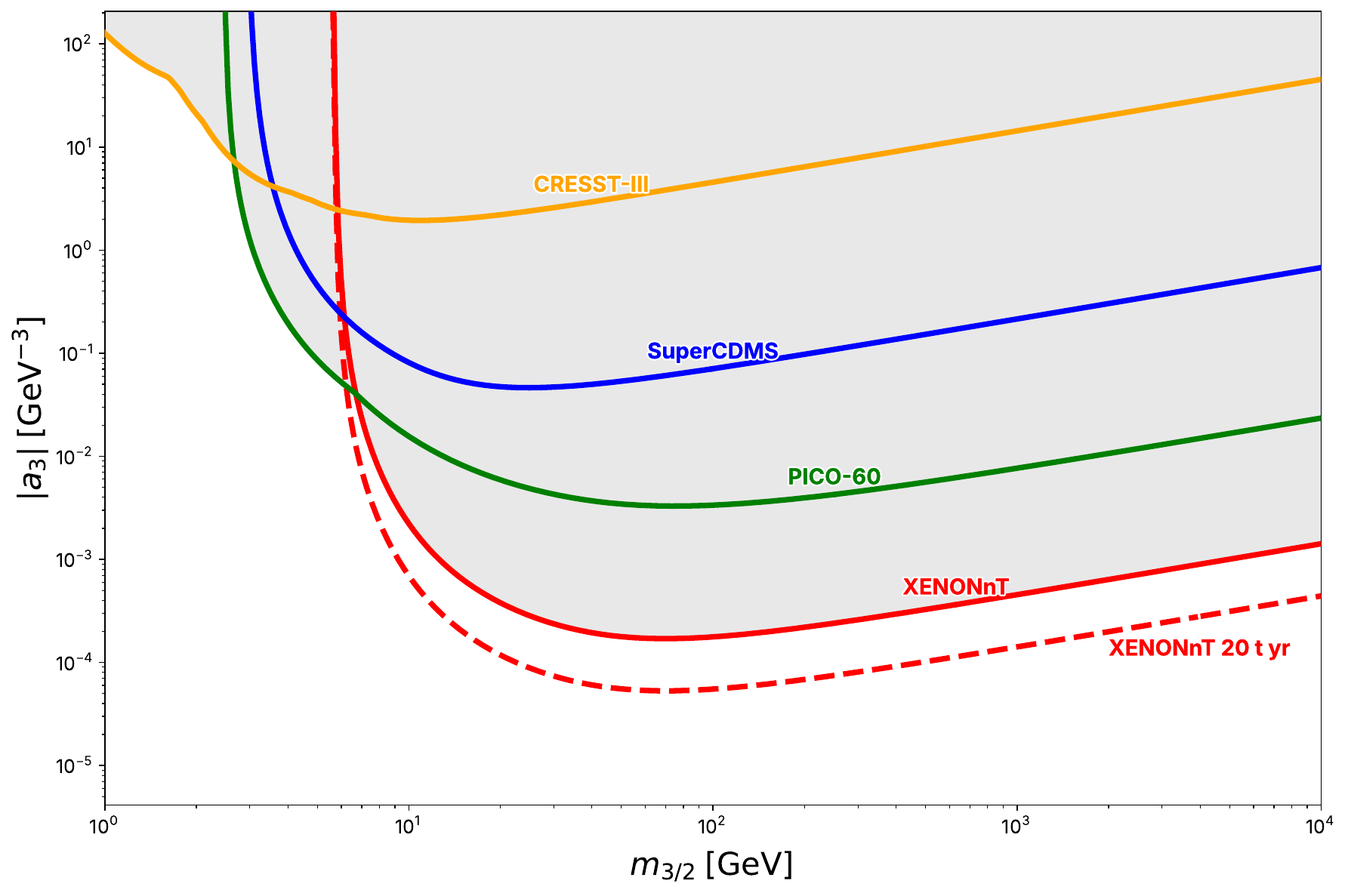}
    \caption{Limits on the three anapole moments $a_1$, $a_2$ and $a_3$. In each subplot we assumed only one moment to be present.}
    \label{fig:limits_model_independent}
\end{figure}

We show in figure~\ref{fig:limits_model_independent} the $90\%$ CL upper limits on the three anapole moments derived from XENONnT~\cite{XENONnT2025}, SuperCDMS~\cite{superCDMS}, PICO-60~\cite{pico60} and CRESST-III~\cite{CRESST:2019axx} data. In each panel, we assume that only one multipole moment is non-vanishing. We follow the nuclear-recoil formalism of ref.~\cite{Hisano:2020qkq}, adapted to the spin-$3/2$ electromagnetic multipoles derived above. To estimate the future sensitivity, we assume a XENONnT-like experiment with an exposure of $\SI{20}{\tonne\year}$. 

\section{Gravitino electromagnetic multipoles}
\label{sec:gravitino}

In this section, we focus on an archetype spin 3/2 Majorana fermion, namely the gravitino in spontaneously broken $N=1$ supergravity~\cite{Freedman:1976xh,Deser:1976eh}. The interaction terms of the gravitino with the different components of the chiral and vector superfields read~\cite{CREMMER1983413}: 
\begin{align}
\mathcal{L}_{\overline{\Psi} \Psi \overline{\chi} \chi}
=&\, \frac{i}{8 M_P^2} \bigg( 
    \overline{\chi}^j_R \gamma^\mu \chi^j_R 
    \left[
        \overline{\Psi}_\mu \gamma_r \sigma^{rs} \Psi_s 
        - \overline{\Psi}_\mu \gamma_s \sigma^{rs} \Psi_r 
        - \overline{\Psi}_r \gamma_\mu \sigma^{rs} \Psi_s 
    \right] \notag \\
&\quad \quad \quad  + \frac{1}{2} \bar{\chi}^j_R \gamma_\delta \chi^j_R
    \left[
        \varepsilon^{\mu \nu \rho \delta} \overline{\Psi}_\mu \gamma_\nu \Psi_\rho 
        - i \overline{\Psi}^\alpha \gamma_5 \gamma^\delta \Psi_\alpha 
    \right] + (\text{L} \leftrightarrow \text{R}) \bigg) + \text{h.c.} \notag \\
\mathcal{L}_{\overline{\Psi} \Psi \overline{\lambda} \lambda}
=&\, \frac{i}{16 M_P^2} \bigg( 
    \overline{\lambda}^A \gamma^\mu \lambda^A 
    \left[
        \overline{\Psi}_\mu \gamma_r \sigma^{rs} \Psi_s 
        - \overline{\Psi}_\mu \gamma_s \sigma^{rs} \Psi_r 
        - \overline{\Psi}_r \gamma_\mu \sigma^{rs} \Psi_s 
    \right] \notag \\
&\quad \quad \quad  + \overline{\lambda}^A \gamma^\mu \sigma^{\rho \sigma} \Psi_\mu 
    \overline{\Psi}_\rho \gamma_\sigma \lambda^A \bigg) + \text{h.c.} \notag \\
\mathcal{L}_{\overline{\Psi} \Psi \phi \phi}
=&\, \frac{1}{8 M_P^2} \bigg( 
    \varepsilon^{\mu \nu \rho \sigma} \overline{\Psi}_\mu \gamma_\nu \Psi_\rho 
    \phi^\dag \partial_\sigma \phi 
    + 4i  m_G \phi^\dag \phi \overline{\Psi}_{L \mu} \sigma^{\nu \mu} \Psi_{R \nu} 
    + (\text{L} \leftrightarrow \text{R}) \bigg) + \text{h.c.} \notag \\
\mathcal{L}_{\overline{\Psi} \phi \gamma \chi}
=&\, \frac{e}{\sqrt{2} M_P} \left( 
    \overline{\Psi}_\mu \gamma^\alpha A_\alpha \phi^{\dag j}_L \gamma^\mu \chi_L^j 
    + (\text{L} \leftrightarrow \text{R}) \right) + \text{h.c.} \notag \\
\mathcal{L}_{\overline{\lambda} \gamma \Psi}
=&\, \frac{1}{8 M_P} \left( 
    \overline{\lambda}^A \gamma^\mu \sigma^{\nu \rho} \Psi_\mu F_{\nu \rho}^A \right) 
    + \text{h.c.} \notag \\
\mathcal{L}_{\overline{\Psi} \chi \phi}
=&\, \left( \frac{i}{\sqrt{2} M_P} 
    \overline{\Psi}_\mu \slashed{\partial} \phi^{\dag j}_L \gamma^\mu \chi_L^j 
    + (\text{L} \leftrightarrow \text{R}) \right) + \text{h.c.} \, ,
\end{align}
where $\Psi_\mu$ is the gravitino field with mass $m_G$, $\chi^j$ denote the spin-1/2 fields, with masses $m_{\chi_j}$,  $\lambda^A$ the gauginos, $F^A_{\mu\nu}$ the field strength tensors, and $\phi_1^j$, $\phi_2^j$ the scalar mass eigenstates, with masses $m_{\phi_{1j}}$ and $m_{\phi_{2j}}$, constructed from mixing the interaction eigenstates $\phi_L^j$, $\phi_R^j$ as
\begin{align}
\phi_1^j &= \phi_L^j \cos\theta_j + \phi_R^j e^{i\varphi}\sin\theta_j, \nonumber \\
\phi_2^j &= -e^{-i\varphi}\phi_L^j \sin\theta_j + \phi_R^j \cos\theta_j.
\end{align}
Here, $\varphi$ is the $CP$-violating angle, which is generated by the supersymmetry breaking terms and contained in the non-diagonal entries of the chiral sfermion mass matrix.  

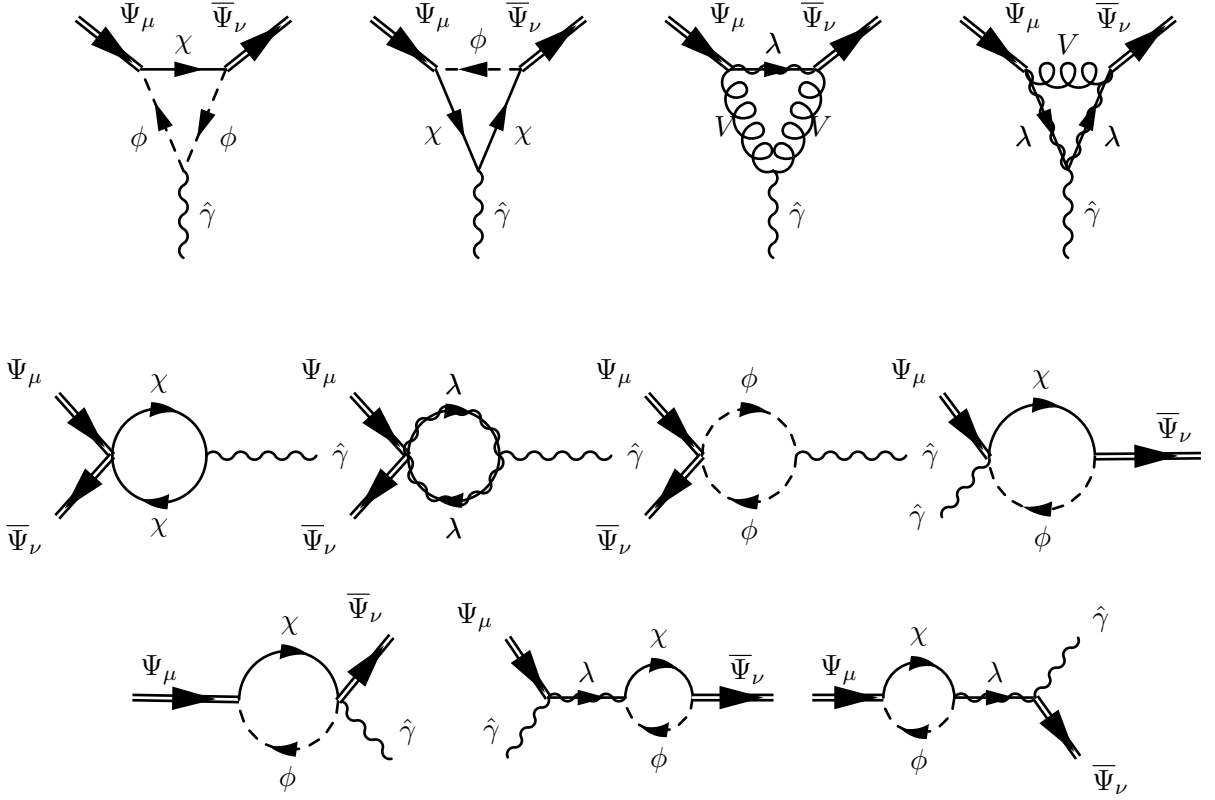
\begin{figure}[t!]
      \centering
 \subfloat{\begin{fmffile}{feyngraph2}
     \begin{fmfgraph*}(100,90)
    \fmfleft{i2,i1}
    \fmfright{o4,o3,o2,o1}
    \fmfbottom{a1,a2,a3}
    \fmf{double_arrow,tension=0.6,lab=$\Psi_\mu$,lab.sid=left}{i1,v1}
    \fmf{double_arrow,tension=.6,lab=$\overline{\Psi}_\nu$,lab.sid=left}{v2,o1}
    \fmf{scalar,tension=.3,lab=$\phi$,lab.sid=left}{v2,v4,v1}
    \fmf{fermion,tension=.3,lab=$\chi$,lab.sid=left}{v1,v2}
    \fmf{photon,tension=0.7,lab=$\hat{\gamma}$,lab.sid=left}{v4,a2}
  \end{fmfgraph*}
\end{fmffile}}
     \subfloat{\begin{fmffile}{feyngraph3}
     \begin{fmfgraph*}(100,90)
    \fmfleft{i2,i1}
    \fmfright{o4,o3,o2,o1}
    \fmfbottom{a1,a2,a3}
    \fmf{double_arrow,tension=0.6,lab=$\Psi_\mu$,lab.sid=left}{i1,v1}
    \fmf{double_arrow,tension=.6,lab=$\overline{\Psi}_\nu$,lab.sid=left}{v2,o1}
    \fmf{fermion,tension=.3,lab=$\chi$,lab.sid=right}{v1,v4,v2}
    \fmf{scalar,tension=.3,lab=$\phi$,lab.sid=right}{v2,v1}
    \fmf{photon,tension=0.7,lab=$\hat{\gamma}$,lab.sid=left}{v4,a2}
  \end{fmfgraph*}
\end{fmffile}}
\subfloat{\begin{fmffile}{feyngraph4}
     \begin{fmfgraph*}(100,90)
    \fmfleft{i2,i1}
    \fmfright{o4,o3,o2,o1}
    \fmfbottom{a1,a2,a3}
    \fmf{double_arrow,tension=0.6,lab=$\Psi_\mu$,lab.sid=left}{i1,v1}
    \fmf{double_arrow,tension=.6,lab=$\overline{\Psi}_\nu$,lab.sid=left}{v2,o1}
    \fmf{gluon,tension=.3,lab=$V$,lab.sid=right}{v1,v4,v2}
    \fmf{fermion,tension=.3,lab=$\lambda$,lab.sid=left}{v1,v2}
    \fmf{photon,tension=.01,lab=$\lambda$,lab.sid=left}{v1,v2}
    \fmf{photon,tension=0.7,lab=$\hat{\gamma}$,lab.sid=left}{v4,a2}
  \end{fmfgraph*}
\end{fmffile}}
\subfloat{\begin{fmffile}{feyngraph5}
     \begin{fmfgraph*}(100,90)
    \fmfleft{i2,i1}
    \fmfright{o4,o3,o2,o1}
    \fmfbottom{a1,a2,a3}
    \fmf{double_arrow,tension=0.6,lab=$\Psi_\mu$,lab.sid=left}{i1,v1}
    \fmf{double_arrow,tension=.6,lab=$\overline{\Psi}_\nu$,lab.sid=left}{v2,o1}
    \fmf{fermion,tension=.3,lab=$\lambda$,lab.sid=right}{v1,v4,v2}
    \fmf{photon,tension=.01,lab=$\lambda$,lab.sid=right}{v1,v4,v2}
    \fmf{gluon,tension=.3,lab=$V$,lab.sid=left}{v1,v2}
    \fmf{photon,tension=0.7,lab=$\hat{\gamma}$,lab.sid=left}{v4,a2}
  \end{fmfgraph*}
\end{fmffile}}
\vspace{1 cm}
\subfloat{\begin{fmffile}{feyngraph8}
    \begin{fmfgraph*}(100,90)
       \fmfleft{i1,i2,i3,i4,i5}
       \fmfright{o1,o2,o3}
       \fmfv{label=$\Psi_\mu$,r.a=-20}{i4}
       \fmfv{label=$\overline{\Psi}_\nu$,r.a=-20}{i2}
       \fmfv{label=$\hat{\gamma}$,l.a=0}{o2}
       \fmf{double_arrow,tension=1.2}{v1,i2}
       \fmf{double_arrow,tension=1.2}{i4,v1}
       \fmf{photon,tension=1.2}{v2,o2}
       \fmf{fermion,left,tension=0.7,label=$\chi$}{v1,v2}
       \fmf{fermion,left,tension=0.7,label=$\chi$}{v2,v1}
    \end{fmfgraph*}
\end{fmffile}}
\subfloat{\begin{fmffile}{feyngraph9}
    \begin{fmfgraph*}(100,90)
       \fmfleft{i1,i2,i3,i4,i5}
       \fmfright{o1,o2,o3}
      \fmfv{label=$\Psi_\mu$,r.a=-20}{i4}
       \fmfv{label=$\overline{\Psi}_\nu$,r.a=-20}{i2}
       \fmfv{label=$\hat{\gamma}$,l.a=0}{o2}
        \fmf{double_arrow,tension=1.2}{v1,i2}
       \fmf{double_arrow,tension=1.2}{i4,v1}
       \fmf{photon,tension=1.2}{v2,o2}
       \fmf{fermion,left,tension=0.7,label=$\lambda$}{v1,v2}
       \fmf{photon,left,tension=0.035,label=$\lambda$}{v1,v2}
       \fmf{fermion,left,tension=0.7,label=$\lambda$}{v2,v1}
        \fmf{photon,left,tension=0.035,label=$\lambda$}{v2,v1}
    \end{fmfgraph*}
\end{fmffile}}
\subfloat{\begin{fmffile}{feyngraph10}
    \begin{fmfgraph*}(100,90)
       \fmfleft{i1,i2,i3,i4,i5}
       \fmfright{o1,o2,o3}
       \fmfv{label=$\Psi_\mu$,r.a=-20}{i4}
       \fmfv{label=$\overline{\Psi}_\nu$,r.a=-20}{i2}
       \fmfv{label=$\hat{\gamma}$,l.a=0}{o2}
       \fmf{double_arrow,tension=1.2}{v1,i2}
       \fmf{double_arrow,tension=1.2}{i4,v1}
       \fmf{photon,tension=1.2}{v2,o2}
       \fmf{scalar,left,tension=0.7,label=$\phi$}{v1,v2}
       \fmf{scalar,left,tension=0.7,label=$\phi$}{v2,v1}
    \end{fmfgraph*}
\end{fmffile}}
\subfloat{\begin{fmffile}{feyngraph6}
    \begin{fmfgraph*}(100,90)
       \fmfleft{i1,i2,i3,i4,i5}
       \fmfright{o1,o2,o3}
       \fmfv{label=$\Psi_\mu$,r.a=0}{i4}
       \fmfv{label=$\overline{\Psi}_\nu$,l.a=120}{o2}
       \fmfv{label=$\hat{\gamma}$,l.a=180}{i2}
       \fmf{double_arrow,tension=1.3}{i4,v1}
       \fmf{photon,tension=1.4}{i2,v1}
       \fmf{double_arrow,tension=1.2}{v2,o2}
       \fmf{fermion,left,tension=0.6,label=$\chi$}{v1,v2}
       \fmf{scalar,left,tension=0.6,label=$\phi$}{v2,v1}
    \end{fmfgraph*}
\end{fmffile}}

\subfloat{\begin{fmffile}{feyngraph7}
    \begin{fmfgraph*}(100,90)
       \fmfleft{i1,i2,i3}
       \fmfright{o1,o2,o3,o4,o5}
       \fmfv{label=$\Psi_\mu$,l.a=60}{i2}
       \fmfv{label=$\overline{\Psi}_\nu$,l.a=120}{o4}
       \fmfv{label=$\hat{\gamma}$,l.a=60}{o2}
       \fmf{double_arrow,tension=1.3}{i2,v1}
       \fmf{photon,tension=1.4}{v2,o2}
       \fmf{double_arrow,tension=1.2}{v2,o4}
       \fmf{fermion,left,tension=0.7,label=$\chi$}{v1,v2}
       \fmf{scalar,left,tension=0.7,label=$\phi$}{v2,v1}
    \end{fmfgraph*}
\end{fmffile}}
\hspace{1 cm}
\begin{fmffile}{feyngraphG1}
  \begin{fmfgraph*}(100,90)
    \fmfstraight
    \fmfleft{i1,i2,i3,i4,i5}
    \fmfright{o1,o2,o3}
    \fmf{photon}{i2,v1}
    \fmf{double_arrow}{i4,v1}
    \fmf{fermion,label=$\lambda$,lab.sid=left}{v1,v2}
    \fmf{photon,tension=0.05}{v1,v2}
     \fmf{fermion,left,tension=0.6,label=$\chi$}{v2,v3}
     \fmf{scalar,left,tension=0.6,label=$\phi$}{v3,v2}
    \fmf{double_arrow,tension=1}{v3,o2}
     \fmfv{label=$\Psi_\mu$,r.a=60}{i4}
     \fmfv{label=$\overline{\Psi}_\nu$,l.a=120}{o2}
     \fmfv{label=$\hat{\gamma}$,l.a=120}{i2}
  \end{fmfgraph*}
\end{fmffile}
     \begin{fmffile}{feyngraphG2}
  \begin{fmfgraph*}(100,90)
    \fmfstraight
    \fmfleft{i1,i2,i3}
    \fmfright{o1,o2,o3,o4,o5}
    \fmf{ double_arrow,tension=1.2}{i2,v1}
       \fmf{fermion,left,tension=0.6,label=$\chi$}{v1,v2}
     \fmf{scalar,left,tension=0.6,label=$\phi$}{v2,v1}
    \fmf{fermion,tension=1,label=$\lambda$,lab.sid=left}{v2,v3}
    \fmf{photon,tension=0.05}{v2,v3}
    \fmf{double_arrow,tension=1}{v3,o2}
    \fmf{photon,tension=1}{v3,o4}
      \fmfv{label=$\Psi_\mu$,l.a=60}{i2}
     \fmfv{label=$\overline{\Psi}_\nu$,r.a=200}{o2}
     \fmfv{label=$\hat{\gamma}$,r.a=60}{o4}
  \end{fmfgraph*}
\end{fmffile}

    \caption{Feynman diagrams contributing to the one loop electromagnetic moments of the gravitino. Diagrams with reversed internal propagators have been omitted.}
    \label{fig:triangleSUGRA}
\end{figure}

These interactions generate a coupling of the gravitino with the photon, through the one loop diagrams shown in figure~\ref{fig:triangleSUGRA} plus the corresponding diagrams  with reversed internal propagators, due to the Majorana nature of the gravitino~\cite{DENNER1992467,DennerMajorana1992}. To ensure the gauge invariance in the calculations, we employed the Background Field Method (BFM) in Feynman gauge, which coincides with the pinch-technique construction and satisfies Ward identities~\cite{bfm1,bfm2,DENNER1994420BFM,bfm4}. In the figure, the background photon is denoted as $\hat{\gamma}$.

To extract the moments after tensorial reduction it is convenient to construct the projectors $\mathcal{P}_i^{\mu\nu\rho}$ such that
\begin{equation}
a_i=\operatorname{Tr}\!\left[\mathcal P_i^{\mu\nu\rho}\Gamma_{\mu\nu\rho}\right]\bigg|_{q^2 \rightarrow 0},\qquad i=1,2 \, ,
\label{eq:Ai_projector_extraction}
\end{equation}
where $\Gamma_{\mu\nu\rho}$ denotes the amplitude without the spin-3/2 spinors. We give explicit form of the relevant projectors $\mathcal{P}_i$ in Appendix~\ref{app:AppendixA}.

We find that most of the diagrams vanish identically. Triangle diagrams involving the vector supermultiplet and fermionic bubble diagrams give no contribution due to an exact cancellation between the corresponding Majorana-conjugate amplitudes. The last four diagrams in figure~\ref{fig:triangleSUGRA} vanish upon imposing the on-shell conditions for the external spin-$3/2$ particles. Consequently, the only nonvanishing contributions to the gravitino anapole moments at one loop arise from triangle diagrams involving the chiral multiplet and from purely scalar bubble diagrams. These contributions generate two non-zero parity-violating moments, while the $CP$-violating moment vanishes at this order.\footnote{This result follows from the structure of the contributing diagrams and the Majorana nature of the external particles. The scalar bubble contributions are diagonal in chirality and therefore do not probe the $CP$-violating phases associated with off-diagonal entries of the chiral mass matrix. In the triangle diagrams, the potentially $CP$-violating contributions arise from chirality-flipping terms, which cancel upon summing the corresponding Majorana-conjugate amplitudes. Thus, despite the presence of potentially complex couplings in the underlying theory, no $CP$-violating electromagnetic moments are generated at one loop.} Our result therefore differs from that of~\cite{no32em}, which concluded that all gravitino electromagnetic multipole moments vanish at one loop.

Concretely, we obtain:\footnote{The Feynman diagrams throughout this work are evaluated with the help of \texttt{Package-X}~\cite{PATEL201766,PATEL2015276} and \texttt{FeynCalc}~\cite{MERTIG1991345,SHTABOVENKO2016432,SHTABOVENKO2020107478}. The Feynman diagrams are drawn with \texttt{Feynmf}~\cite{Ohl:1995kr}. }
\begin{equation}
\begin{aligned}
     a_1^\Psi&= \sum_{j}\dfrac{ N_{c,j} e Q_{j} \cos{2 \theta_j}}{ 192 \pi^2 M_P^2 }\left[F_1(m_{\chi_j}^2/m_G^2,m_{\phi_{1j}}^2/m_G^2)-F_1(m_{\chi_j}^2/m_G^2,m_{\phi_{2j}}^2/m_G^2)\right], \\
      a_2^\Psi&= \sum_{j}\dfrac{ N_{c,j} e Q_{j} \cos{2 \theta_j}}{ 48 \pi^2 M_P^2 m_G }\left[F_2(m_{\chi_j}^2/m_G^2,m_{\phi_{1j}}^2/m_G^2)-F_2(m_{\chi_j}^2/m_G^2,m_{\phi_{2j}}^2/m_G^2)\right],
\end{aligned}
\end{equation}
where the sum runs over all fermions in the loop, with charge $e Q_j$ and color factor $N_{c,j}$. Further, $F_i$ are  loop functions defined as:
\begin{align}
  F_1(\xi^2,\eta^2)&=4+6(\eta^2-\xi^2)-\ln \eta^2+\big[3(\xi^2-\eta^2)^2-4\xi^2-2\eta^2-1\big]\ln\dfrac{ \xi^2}{\eta^2}\nonumber\\
  &\quad+2\,(1-3\xi^2+3\eta^2)\,\lambda(\xi^2,\eta^2)\,f(\xi^2,\eta^2),\label{eq:F1}\\
  F_2(\xi^2,\eta^2)&=1-2\xi^2+2\eta^2+\Big[(\xi^2-\eta^2)^2-\xi^2+\tfrac12\Big]\ln\dfrac{ \xi^2}{\eta^2}\nonumber\\
  &\quad+\Big[2(\eta^2-\xi^2)^3-2(\eta^2-\xi^2)(2\xi^2+\eta^2)+\eta^2-3\xi^2+\tfrac35\Big]f(\xi^2,\eta^2),
  \label{eq:F2}
\end{align}
with
\begin{equation}
  f(\xi^2,\eta^2)=
  \begin{cases}
  \dfrac{1}{2\sqrt{\lambda}}\ln\dfrac{\xi^2+\eta^2-1+\sqrt{\lambda}}{\xi^2+\eta^2-1-\sqrt{\lambda}}, & \lambda>0,\\
  \dfrac{1}{\sqrt{-\lambda}}\Big[\arctan\dfrac{1+\xi^2-\eta^2}{\sqrt{-\lambda}}+\arctan\dfrac{1+\eta^2-\xi^2}{\sqrt{-\lambda}}\Big], & \lambda<0,\\
  \dfrac{1}{\xi^2+\eta^2-1}, & \lambda=0 ,
  \end{cases}
  \label{eq:f}
\end{equation}
and $\lambda(x,y)=\lambda(1,x,y)=\lambda=1+x^2+y^2-2xy-2x-2y$ is the dimensionless Källén function. Here it must be stressed that left and right handed sectors separately are divergent, but the divergence is independent of the chiral sector and cancels exactly in the summation. Furthermore, as expected, the anapole moments vanish in the limit of exact supersymmetry.

For the specific limit where the scalars are much heavier than the fermions, $m_{\phi_{1j}},m_{\phi_{2j}} \gg m_G, m_{\chi_j}$,
\begin{align}
  a_1^\Psi&\simeq\sum_{j}\frac{eN_{c,j}Q_j\cos2\theta_j}{192\pi^2M_P^2}\ln\frac{m_{\phi_{2j}}^2}{m_{\phi_{1j}}^2},\\
  a_2^\Psi&\simeq-\sum_{j}\frac{e\,m_G N_{c,j}Q_j\cos2\theta_j}{720\pi^2M_P^2}\Bigg[\frac{25+12\ln(m_{\chi_j}^2/m_{\phi_{1j}}^2)}{m_{\phi_{1j}}^2}-\frac{25+12\ln(m_{\chi_j}^2/m_{\phi_{2j}}^2)}{m_{\phi_{2j}}^2}\Bigg].
\end{align}

On the other hand, if the spin-1/2 fermions are heavier than the scalars and the gravitino,  $m_{\chi_j} \gg m_G, m_{\phi_{1j}},m_{\phi_{2j}}$,
\begin{align}
  a_1^\Psi&\simeq\sum_{j}\frac{eN_{c,j}Q_j\cos2\theta_j}{192\pi^2M_P^2}\Big[\dfrac{1}{2}\ln\frac{m_{\phi_{1j}}^2}{m_{\phi_{2j}}^2}+\frac{4m_{\phi_{1j}}^2\ln(m_{\chi_j}^2/m_{\phi_{1j}}^2)-6m_{\phi_{1j}}^2}{m_{\chi_j}^2}-(1\leftrightarrow2)\Big],\\
  a_2^\Psi&\simeq\sum_{j}\frac{e\,m_G N_{c,j}Q_j\cos2\theta_j}{240\pi^2M_P^2}\frac{1}{m_{\chi_j}^2}\ln\frac{m_{\phi_{1j}}^2}{m_{\phi_{2j}}^2}.
\end{align}

From a theoretical perspective, it is remarkable that the gravitino can possess non-zero electromagnetic multipole moments. However, these are parametrically suppressed by powers of the Planck mass, making their experimental detection extremely challenging.

\section{Electromagnetic multipoles in a simplified model with spin-3/2 Majorana fermions}
\label{sec:simplified-model}

We consider now a simplified model in which a spin-$3/2$ Majorana fermion $\Psi_\mu$ interacts with an electrically charged spin-$1/2$ fermion $\lambda$, of mass $m_\lambda$ and charge $-e$, and a complex massive vector $V_\mu$, of mass $m_V$ and charge $+e$. We regard this model as an effective field theory (EFT), rather than as a complete theory valid at arbitrarily high energies. In particular, generic interactions of a massive Proca vector may lead to scattering amplitudes involving longitudinal polarizations that grow with energy, eventually invalidating perturbation theory. We therefore assume that the effective description is applicable below a cut-off, with the effects of the ultraviolet completion encoded in local operators. We do not specify the underlying dynamics, which could involve a spontaneously broken gauge theory or composite particles. The free spin-$3/2$ field is described by the Rarita-Schwinger Lagrangian eq.~(\ref{eq:RS}), while the free Dirac and Proca Lagrangians take the form
\begin{align}
\mathcal L_\lambda&=\bar\lambda(i\gamma^\mu\partial_\mu-m_\lambda)\lambda\,, \nonumber \\
\mathcal L_P&=-\frac12 V_{\mu\nu}^\dagger V^{\mu\nu}+m_V^2V_\mu^\dagger V^\mu\,,
\label{eq:lambda-free}
\end{align}
where $V_{\mu\nu}=\partial_\mu V_\nu-\partial_\nu V_\mu$.

The electromagnetic interaction of the charged fermion and the dipole-type coupling of the complex vector are written as
\begin{align}
\mathcal L_{\lambda\gamma}&=-e\bar\lambda\gamma^\mu A_\mu\lambda\,,\\
\mathcal L_{V\gamma}&=ie A^\mu\left(V^\dag_{\mu\nu}V^\nu - V^{\dag\nu}V_{\mu\nu}\right) -ie\kappa_VV_\mu^\dagger V_\nu F^{\mu\nu} \nonumber \\
& \quad + e^2 \left(A_\mu A_\nu V^{\dag\mu} V^\nu-A_\mu A^\mu V^\dag_\nu V^\nu\right)\,,
\end{align}
where $A_\mu$ is the electromagnetic potential and $F_{\mu\nu}$ its field strength. The parameter $\kappa_V$ specifies the nonminimal magnetic coupling of the charged vector, in addition to the interactions associated with electromagnetic covariant derivatives. In the usual convention $\kappa_V=0$ corresponds to minimal Proca coupling, whereas $\kappa_V=1$ gives the gyromagnetic ratio $g_V=2$ characteristic of Yang-Mills gauge bosons~\cite{HAGIWARA1987253,FerraraPorratiTelegdi1992}. We allow either choice without imposing a specific ultraviolet completion. If $V_\mu$ is composite, its constituents may generate additional electromagnetic interactions not captured by the pointlike description.

The interaction of the spin-$3/2$ Majorana fermion with the charged fermion and vector is
\begin{equation}\label{eq:spin32-portal}
\mathcal L_{\mathrm{int}}=\bar\Psi_\mu(v_LP_L+v_RP_R)V^{+\mu}\lambda+\mathrm{h.c.}\,,
\end{equation}
where $v_L$ and $v_R$ can be complex. The interaction preserves parity for $v_L=\pm v_R$, charge conjugation for $|v_L|=|v_R|$, and $CP$ for $\operatorname{Im}(v_Lv_R^*)=0$.

\begin{figure}[t!]
      \centering
 
\subfloat{\begin{fmffile}{feyngraph4s}
     \begin{fmfgraph*}(130,110)
    \fmfleft{i2,i1}
    \fmfright{o4,o3,o2,o1}
    \fmfbottom{a1,a2,a3}
    \fmf{double_arrow,tension=0.6,lab=$\Psi_\mu$,lab.sid=left}{i1,v1}
    \fmf{double_arrow,tension=.6,lab=$\overline{\Psi}_\nu$,lab.sid=left}{v2,o1}
    \fmf{gluon,tension=.3,lab=$V$,lab.sid=right}{v1,v4,v2}
    \fmf{fermion,tension=.3,lab=$\lambda$,lab.sid=left}{v1,v2}
    \fmf{photon,tension=.01,lab=$\lambda$,lab.sid=left}{v1,v2}
    \fmf{photon,tension=0.7,lab=$\gamma$,lab.sid=left}{v4,a2}
  \end{fmfgraph*}
\end{fmffile}}
\subfloat{\begin{fmffile}{feyngraph5s}
     \begin{fmfgraph*}(130,110)
    \fmfleft{i2,i1}
    \fmfright{o4,o3,o2,o1}
    \fmfbottom{a1,a2,a3}
    \fmf{double_arrow,tension=0.6,lab=$\Psi_\mu$,lab.sid=left}{i1,v1}
    \fmf{double_arrow,tension=.6,lab=$\overline{\Psi}_\nu$,lab.sid=left}{v2,o1}
    \fmf{fermion,tension=.3,lab=$\lambda$,lab.sid=right}{v1,v4,v2}
    \fmf{photon,tension=.01,lab=$\lambda$,lab.sid=right}{v1,v4,v2}
    \fmf{gluon,tension=.3,lab=$V$,lab.sid=left}{v1,v2}
    \fmf{photon,tension=0.7,lab=$\gamma$,lab.sid=left}{v4,a2}
  \end{fmfgraph*}
\end{fmffile}}

    \caption{Feynman diagrams generating an anapole moment for the spin-3/2 Majorana fermion at the one-loop level in the simplified model  described in section~\ref{sec:simplified-model}.}
    \label{fig:triangleProca}
\end{figure}
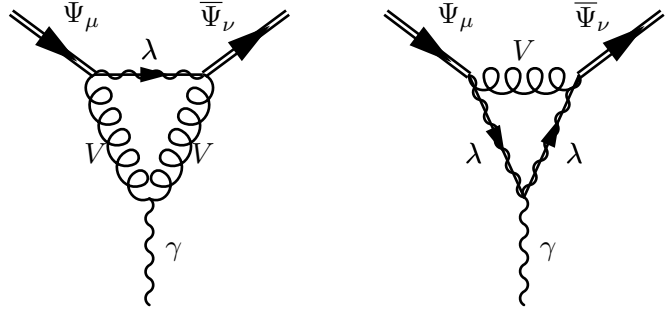

The electromagnetic interactions of $\lambda$ and $V_\mu$ induce anapole moments of $\Psi_\mu$ at one loop through the diagrams in figure~\ref{fig:triangleProca}. We find that the $C$- and $P$-violating moments $a_1$ and $a_2$ are in general non-zero. The longitudinal part of the Proca propagator leads to a stronger ultraviolet sensitivity than in models with scalar mediators, namely $a_1$ contains a UV divergence, whereas $a_2$ is finite at this order. The divergent part of $a_1$ reads
\begin{equation}\label{eq:a1-uv}
 a_1^{\mathrm{UV}}=\frac{e(|v_L|^2-|v_R|^2)}{16\pi^2}\frac{1}{\bar\epsilon}\left[\frac{1}{6m_V^2}+\frac{\kappa_V}{48m_V^4}\big(m_{3/2}^2-2m_\lambda^2-8m_V^2\big)\right],
\end{equation}
where $1/\bar\epsilon=1/\epsilon-\gamma_E+\ln(4\pi)$. In the EFT, this divergence is absorbed into the coefficient of the corresponding dimension-six electromagnetic anapole operator. Using dimensional regularization and the $\overline{\mathrm{MS}}$ scheme, the renormalized moment is defined as
\begin{equation}\label{eq:a1-ren}
 a_1^{\mathrm{ren}}(\mu)=C_6(\mu)+a_1^{\lambda V}(\mu)\,,
\end{equation}
where $C_6(\mu)$ denotes the renormalized coefficient of the local operator. Its renormalization-group equation is
\begin{align}
 \frac{dC_6}{d\ln\mu}&=-2\frac{e(|v_L|^2-|v_R|^2)}{16\pi^2}\left[\frac{1}{6m_V^2}+\frac{\kappa_V}{48m_V^4}\big(m_{3/2}^2-2m_\lambda^2-8m_V^2\big)\right]\label{eq:c6-rge-1}\\
 &=-2\frac{e(|v_L|^2-|v_R|^2)}{16\pi^2m_{3/2}^2}K\!\left(\frac{m_\lambda^2}{m_{3/2}^2},\frac{m_V^2}{m_{3/2}^2};\kappa_V\right),\label{eq:c6-rge-2}
\end{align}
with
\begin{equation}\label{eq:def-K}
 K(x,y;\kappa_V)\equiv\frac{1}{6y}+\kappa_V\frac{1-2x-8y}{48y^2}\,.
\end{equation}

We introduce a matching scale $\Lambda$, at or below the physical cutoff of the EFT, at which the unknown ultraviolet dynamics is encoded in the coefficients of local operators. Integrating eq.~\eqref{eq:c6-rge-2} between $\Lambda$ and $\mu$, we obtain
\begin{equation}\label{eq:c6-running}
 C_6(\mu)=C_6(\Lambda)+\frac{e(|v_L|^2-|v_R|^2)}{16\pi^2m_{3/2}^2}K\!\left(\frac{m_\lambda^2}{m_{3/2}^2},\frac{m_V^2}{m_{3/2}^2};\kappa_V\right)\ln\frac{\Lambda^2}{\mu^2}\,.
\end{equation}

Besides, the finite loop contributions read:
\begin{align}
 a_1^{\lambda V}(\mu)&=\frac{e(|v_L|^2-|v_R|^2)}{16\pi^2m_{3/2}^2}\left[G_1\!\left(\frac{m_\lambda^2}{m_{3/2}^2},\frac{m_V^2}{m_{3/2}^2};\kappa_V\right)+K\!\left(\frac{m_\lambda^2}{m_{3/2}^2},\frac{m_V^2}{m_{3/2}^2};\kappa_V\right)\ln\frac{\mu^2}{m_V^2}\right],\label{eq:a1-loop}\\
 a_2^{\lambda V}&=\frac{e(|v_L|^2-|v_R|^2)}{16\pi^2m_{3/2}^3}G_2\!\left(\frac{m_\lambda^2}{m_{3/2}^2},\frac{m_V^2}{m_{3/2}^2};\kappa_V\right).\label{eq:a2-loop}
\end{align}

Finally, combining the renormalized loop contribution and the running of the local coefficient, the physical static moments are
\begin{align}
 a_1&=C_6(\Lambda)+\frac{e(|v_L|^2-|v_R|^2)}{16\pi^2m_{3/2}^2}\left[G_1\!\left(\frac{m_\lambda^2}{m_{3/2}^2},\frac{m_V^2}{m_{3/2}^2};\kappa_V\right)+K\!\left(\frac{m_\lambda^2}{m_{3/2}^2},\frac{m_V^2}{m_{3/2}^2};\kappa_V\right)\ln\frac{\Lambda^2}{m_V^2}\right],\label{eq:a1-full}\\
 a_2&=C_7(\Lambda)+\frac{e(|v_L|^2-|v_R|^2)}{16\pi^2m_{3/2}^3}G_2\!\left(\frac{m_\lambda^2}{m_{3/2}^2},\frac{m_V^2}{m_{3/2}^2};\kappa_V\right),\label{eq:a2-full}
\end{align}
where the functions $G_1$ and $G_2$ are given in Appendix~\ref{app:AppendixB}. The dependence on the intermediate renormalization scale $\mu$ cancels in the physical moment. The matching coefficients  $C_6(\Lambda)$ and $C_7(\Lambda)$ cannot be fixed by the low-energy calculation alone. In particular, even though $a_2$ is finite at one loop, it may receive an independent contribution from ultraviolet physics. For a composite spin-$3/2$ particle, these terms can also describe intrinsic electromagnetic structure associated with the confining dynamics. For simplicity, we will set them to zero, therefore our calculation provides a lower limit on the value of the anapole moments (barring cancellations).

It is useful to expand the loop-induced moments in several limiting mass hierarchies. First, for $m_{3/2}\ll m_V,m_\lambda$, we obtain
\begin{align}
 a_1&\simeq\frac{e(|v_L|^2-|v_R|^2)}{1152\pi^2m_V^2(m_\lambda^2-m_V^2)^3}\notag\\&\quad\times\Bigg\{8m_\lambda^6+9m_\lambda^4m_V^2-36m_\lambda^2m_V^4+19m_V^6-6\left(2m_\lambda^6-3m_\lambda^4m_V^2+6m_\lambda^2m_V^4-4m_V^6\right)\ln\frac{m_\lambda^2}{m_V^2}\notag\\
 &\qquad+\frac{\kappa_V}{2m_V^2}\Bigg[6m_\lambda^6(m_\lambda^2+m_V^2)\ln\frac{m_\lambda^2}{m_V^2}-11m_\lambda^8-11m_\lambda^6m_V^2+63m_\lambda^4m_V^4-61m_\lambda^2m_V^6+20m_V^8\Bigg]\notag\\
 &\qquad+3\frac{4m_V^2-\kappa_V(m_\lambda^2+4m_V^2)}{m_V^2}(m_\lambda^2-m_V^2)^3\ln\frac{\Lambda^2}{m_V^2}\Bigg\},  \\
 a_2&\simeq\frac{e(|v_L|^2-|v_R|^2)m_{3/2}}{5760\pi^2m_V^2(m_\lambda^2-m_V^2)^5}\Bigg\{25m_\lambda^8-150m_\lambda^6m_V^2+270m_\lambda^4m_V^4\notag\\
 &\quad-250m_\lambda^2m_V^6+105m_V^8-12\left(m_\lambda^8-4m_\lambda^6m_V^2+m_\lambda^4m_V^4+m_\lambda^2m_V^6-4m_V^8\right)\ln\frac{m_\lambda^2}{m_V^2}\notag\\
 &\quad+5\kappa_V\Bigg[47m_\lambda^8-106m_\lambda^6m_V^2+90m_\lambda^4m_V^4-38m_\lambda^2m_V^6+7m_V^8-12m_\lambda^6(2m_\lambda^2-m_V^2)\ln\frac{m_\lambda^2}{m_V^2}\Bigg]\Bigg\}.\label{eq:light-a2}
\end{align}

As expected, the one-loop contribution to the first and second anapole form factors vanishes when $|v_L|^2-|v_R|^2=0$, which requires $C$ and/or $P$ conservation. Conversely, the generation of these two  anapole moments requires both $C $  and $P$ violation. We also note that at leading order, $a_1$ is independent of the  mass of the spin-3/2 fermion, whereas $a_2$ is linear in $m_{3/2}$. 

For a vector much heavier than both the spin-$3/2$ particle and the charged fermion, $m_V\gg m_{3/2},m_\lambda$, we find
\begin{align}
 a_1&\simeq\frac{e(|v_L|^2-|v_R|^2)}{16\pi^2m_V^2}\left[\frac13\ln\frac{m_V^2}{m_\lambda^2}+\frac{1-\kappa_V}{6}\ln\frac{\Lambda^2}{m_V^2}-\frac{19}{72}-\frac{5\kappa_V}{36}\right],\notag\\
 a_2&\simeq\frac{e(|v_L|^2-|v_R|^2)m_{3/2}}{16\pi^2m_V^4}\left[\frac{2}{15}\ln\frac{m_V^2}{m_\lambda^2}-\frac{7(3+\kappa_V)}{72}\right].
 \label{eq:heavy-vector}
\end{align}

Finally, for $m_\lambda\gg m_{3/2},m_V$, the leading terms are
\begin{align}
 a_1&\simeq\frac{e(|v_L|^2-|v_R|^2)}{16\pi^2}\Bigg\{-\kappa_V\frac{m_\lambda^2}{m_V^4}\left(\frac{1}{24}\ln\frac{\Lambda^2}{m_\lambda^2}+\frac{11}{144}\right)\notag\\
 &\qquad+\frac{1}{m_V^2}\left[\frac{1-\kappa_V}{6}\ln\frac{\Lambda^2}{m_\lambda^2}+\frac{4-11\kappa_V}{36}\right]+\kappa_V\frac{m_{3/2}^2}{m_V^4}\left(\frac{1}{48}\ln\frac{\Lambda^2}{m_\lambda^2}+\frac{13}{576}\right)\Bigg\}, \notag\\
 a_2&\simeq\frac{e(|v_L|^2-|v_R|^2)m_{3/2}}{16\pi^2m_V^2m_\lambda^2}\left[-\left(\frac{1}{30}+\frac{\kappa_V}{3}\right)\ln\frac{m_\lambda^2}{m_V^2}+\frac{5}{72}+\frac{47\kappa_V}{72}\right].
\end{align}

\begin{figure}[t!]
    \centering
 \includegraphics[width=.99\textwidth]{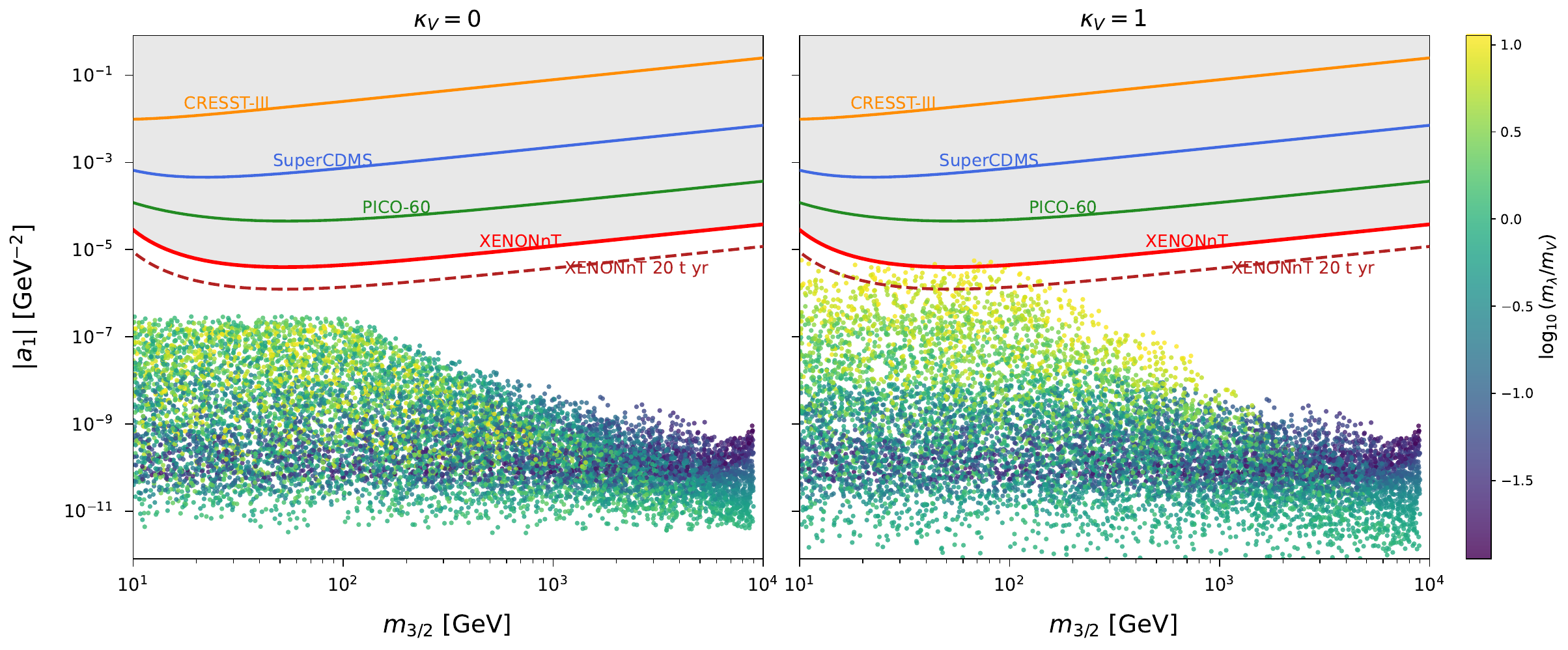}
   \includegraphics[width=.99\textwidth]{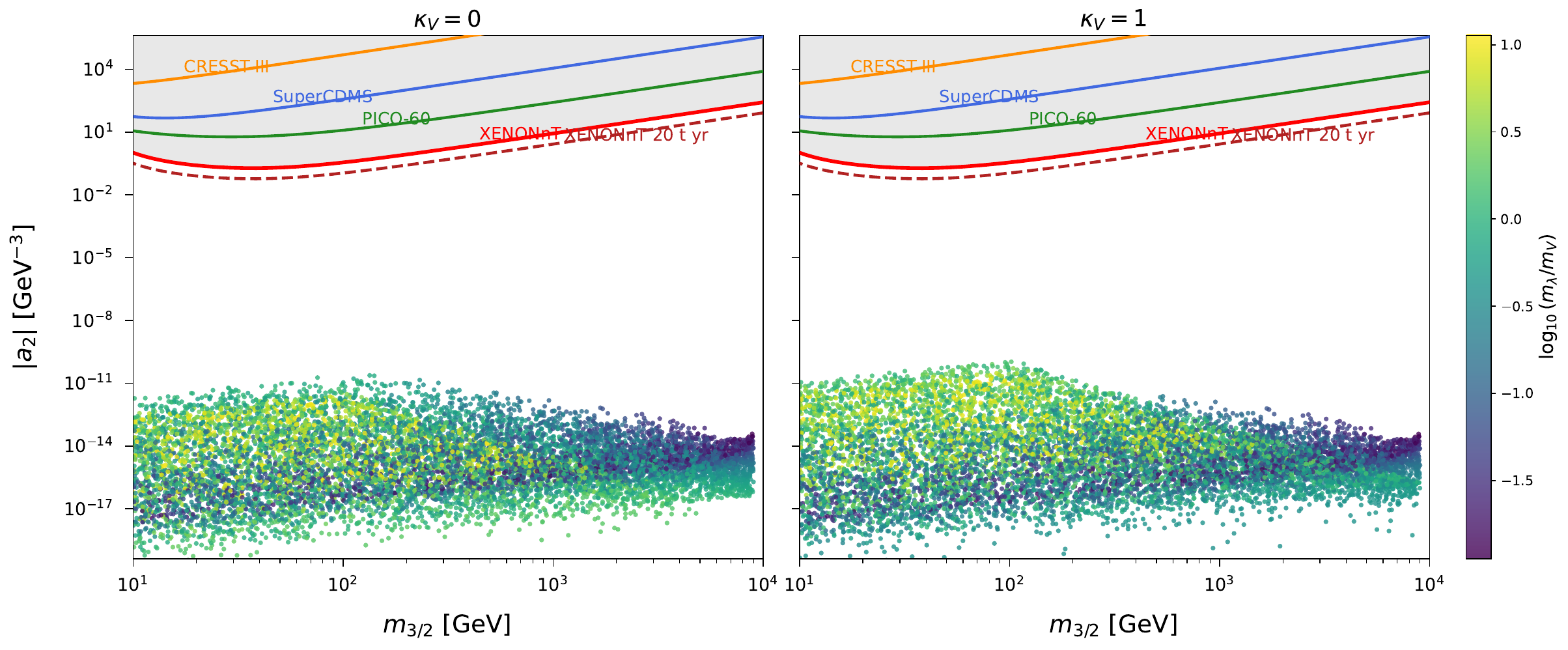}
    \caption{Scatter plot for the radiatively generated anapole moments $|a_1|$ (top) and $|a_2|$ (bottom) in the Proca portal
benchmark, compared with the direct-detection limits. The left and right panels correspond to $\kappa_V=0$ and $\kappa_V=1$,
respectively. }
    \label{fig:proca-scatter}
\end{figure}

\vspace{5cm}

Figure~\ref{fig:proca-scatter} shows the expected values of $a_1$ and $a_2$ for $\kappa_V=0$ (left) and $\kappa_V=1$ (right) under the assumption that $C_6(\Lambda)=C_7(\Lambda)=0$, so that the multipole moments arise entirely from the loop contributions of the charged fermion and vector. We choose $\Lambda=10\,\mathrm{TeV}$ as a reference matching scale and require the particle masses to be smaller than this value.  We scan the region $10\,\mathrm{GeV}\leq m_{3/2}<\Lambda$, $100\,\mathrm{GeV}\leq m_V,m_\lambda\leq\Lambda$, imposing $m_V>m_{3/2}$, so that the two-body decay $\Psi\to V\lambda$ is kinematically closed, and  $\sqrt{|v_L|^2+|v_R|^2}\,m_\lambda/m_V\lesssim4\pi$ to ensure the perturbativity of the couplings.

At low dark-matter masses, $a_1$ exhibits a plateau consistent with the expansion in $m_{3/2}/m_V$, whereas $a_2$ grows approximately linearly with $m_{3/2}$. The upper envelope of $a_1$ is substantially larger in the $\kappa_V=1$ case, where the longitudinal Proca contribution enhances the loop amplitude for $m_\lambda/m_V\gg1$. Destructive interference between different terms can also produce pronounced suppressions near cancellation surfaces. At larger $m_{3/2}$, the requirement $m_V>m_{3/2}$ tends to suppress both moments. For $\kappa_V=1$, the scan reaches $|a_1|\sim5\times10^{-6}\,\mathrm{GeV^{-2}}$, which could be at the reach of current experiments.
\FloatBarrier

\section{Conclusions}
\label{sec:conclusions}

We have investigated the electromagnetic properties of Majorana spin-3/2 dark matter particles and their implications for direct detection experiments. We have first considered the general electromagnetic vertex, parametrized in terms of three form factors, and derived constraints on the corresponding static moments from the non-observation of nuclear recoils. 

We have then calculated the electromagnetic moments of the gravitino in the Minimal Supersymmetric Standard Model coupled to supergravity. We find that both parity-violating anapole moments are generated at one loop, through diagrams involving chiral supermultiplets and scalar loops, whereas the $CP$-violating moment vanishes at this order.  Although the gravitino develops non-vanishing electromagnetic moments, their magnitude is suppressed by the Planck scale, rendering their detection through nuclear recoils far beyond the reach of current and foreseeable experiments.

Finally, we have considered a simplified model where a composite Majorana spin-3/2 particle interacts with a charged fermion and a massive charged vector boson. We find that the two parity-violating anapole moments are generated at one loop. While the second moment is finite, the first requires renormalization and depends on the matching conditions at the ultraviolet scale. We find that the first anapole moment can reach values accessible to future direct detection experiments, particularly for dark matter masses around the electroweak scale and relatively light charged mediators. The prospects are especially interesting when the charged vector has a gyromagnetic ratio $g_V=2$, which can enhance the anapole moment through its longitudinal interactions.

\acknowledgments
We sincerely thank Vladimir Pascalutsa for valuable discussions on the interactions of spin-3/2 fermions, and Merlin Reichard for insightful discussions on electromagnetic multipoles. This work has been supported by the Collaborative Research Center SFB1258 and by the Deutsche Forschungsgemeinschaft (DFG, German Research Foundation) under Germany's Excellence Strategy - EXC-2094 - 390783311.

\appendix

\section{Form factor projectors}\label{app:AppendixA}

In this appendix we present the projectors used to extract the form factors. They are
applied to the renormalized vertex $\Gamma^{\mu\nu\rho}$.  The
four-dimensional spin sum of the spin-3/2 particle is,
\begin{equation}
S_{\alpha\beta}(r)=-(\slashed r+m_{3/2})\left[g_{\alpha\beta}-\frac{\gamma_\alpha\gamma_\beta}{3}
-\frac{2r_\alpha r_\beta}{3m_{3/2}^2}+\frac{r_\alpha\gamma_\beta-r_\beta\gamma_\alpha}{3m_{3/2}}\right],
\label{eq:RS_spin_sum}
\end{equation}
and the projector onto $A_1$ is
\begin{align}
\mathcal P_1^{\mu\nu\rho}={}&
\frac{3m_{3/2}^3}{4q^6(q^2-4m_{3/2}^2)^3}\,
S^{\mu}{}_{\mu'}(p)\Bigg\{
5m_{3/2}q^2\,g^{\nu'\mu'}\left(q^2\gamma^\rho-q^\rho\slashed q\right)\gamma_5\nonumber\\
&-4\left(m_{3/2}^2+q^2\right)\gamma_5
\left[q^{\mu'}q^{\nu'}q^\rho-\frac{q^2}{2}\left(g^{\rho\nu'}q^{\mu'}+g^{\mu'\rho}q^{\nu'}\right)\right]
\Bigg\}S_{\nu'}{}^{\nu}(l),
\label{eq:P1_explicit}
\end{align}
onto $A_2$
\begin{align}
\mathcal P_2^{\mu\nu\rho}={}&
\frac{3m_{3/2}^2}{2q^8(q^2-4m_{3/2}^2)^3}\,
S^{\mu}{}_{\mu'}(p)\Bigg\{
2m_{3/2}q^2\left(m_{3/2}^2+q^2\right)g^{\nu'\mu'}\left(q^2\gamma^\rho-q^\rho\slashed q\right)\gamma_5\nonumber\\
&-4\left(10m_{3/2}^4-4m_{3/2}^2q^2+q^4\right)\gamma_5
\left[q^{\mu'}q^{\nu'}q^\rho-\frac{q^2}{2}\left(g^{\rho\nu'}q^{\mu'}+g^{\mu'\rho}q^{\nu'}\right)\right]
\Bigg\}S_{\nu'}{}^{\nu}(l),
\label{eq:P2_explicit}
\end{align}
and onto $A_3$
\begin{equation}
\mathcal P_3^{\mu\nu\rho}=
\frac{6m_{3/2}^2}{q^6(q^2-4m_{3/2}^2)^2}\,
S^{\mu}{}_{\mu'}(p)
\left[q^{\mu'}q^{\nu'}q^\rho-\frac{q^2}{2}\left(g^{\rho\nu'}q^{\mu'}+g^{\mu'\rho}q^{\nu'}\right)\right]
S_{\nu'}{}^{\nu}(l).
\label{eq:P3_explicit}
\end{equation}
The form factors then follow as
\begin{equation}
A_1(q^2)=\operatorname{Tr}\big[\mathcal P_1^{\mu\nu\rho}\Gamma_{\mu\nu\rho}\big],\qquad
A_2(q^2)=\operatorname{Tr}\big[\mathcal P_2^{\mu\nu\rho}\Gamma_{\mu\nu\rho}\big],\qquad
A_3(q^2)=\operatorname{Tr}\big[\mathcal P_3^{\mu\nu\rho}\Gamma_{\mu\nu\rho}\big].
\end{equation}

\section{Static coefficients}
\label{app:AppendixB}
Here we define the functions $G_1$ and $G_2$ for the moments in the simplified toy model.
\begin{align}
  G_1(x^2,y^2;\kappa_V)&=\frac1{72y^2}\left\{
 25+3y^2-15x^2+6\delta^2
 +3[-2+7y^2-4x^2-3x^2\delta+\delta^3]\ln\frac{x^2}{y^2}\right.\nonumber\\
 &\left.\hspace{0mm}+6[2-y^2-6x^2+7y^4-10x^2y^2+7x^4
 -(y^2+4x^2)\delta^2+\delta^4]f(x^2,y^2)\right\}\nonumber\\
 &+\frac{\kappa_V}{288y^4}\left\{
 16-110y^2-26x^2+81y^4-24x^2y^2-9x^4-(30y^2-6x^2)\delta^2\right.\nonumber\\
 &+3[8y^2+2x^2-18y^4-1+(16y^4-2x^2y^2-2x^4)\delta
 -(5y^2-x^2)\delta^3]\ln\frac{x^2}{y^2}\nonumber\\
 &\left. +6\lambda(x^2,y^2)[1-7y^2-x^2+11y^4-6x^2y^2-x^4
 -(5y^2-x^2)\delta^2]f(x^2,y^2)\right\},\label{eq:finiteF1} \\
 G_2(x^2,y^2;\kappa_V)&=\frac1{180y^2}\left\{
 5[8+3y^2-9x^2+6\delta^2]
 +15[1+y^2-2x^2-2x^2\delta+\delta^3]\ln\frac{x^2}{y^2}\right.\nonumber\\
 &\left.\hspace{0mm}+6[3-15x^2+5y^4-15x^2y^2+20x^4
 -(5y^2+15x^2)\delta^2+5\delta^4]f(x^2,y^2)\right\}\nonumber\\
 &+\frac{\kappa_V}{36y^2}\left\{
 21y^2-15x^2-6\delta^2-20
 +3[2-5y^2+2x^2+(4y^2-2x^2)\delta-\delta^3]\ln\frac{x^2}{y^2}\right.\nonumber\\
 &\left.\hspace{0mm}+6[-2+7y^2-9y^4+7x^2y^2
 +(5y^2-x^2)\delta^2-\delta^4]f(x^2,y^2)\right\},
 \label{eq:finiteF2}
\end{align}
where $\lambda$ is the dimensionless Källén function and $\delta=y^2-x^2$. The function $f(x^2,y^2)$ is defined in the eq.~\ref{eq:f}.

\bibliographystyle{JHEP}
\bibliography{biblio.bib}

\end{document}